# Analysis of hierarchical organization in gene expression networks reveals underlying principles of collective tumor cell dissemination and metastatic aggressiveness of inflammatory breast cancer

Shubham Tripathi[1, 2], Mohit Kumar Jolly[2], Wendy A. Woodward[3, 4], Herbert Levine[2, 5, 6], and Michael W. Deem[2, 5, 6, *]

[1] PhD Program in Systems, Synthetic, and Physical Biology, Rice University, Houston, TX 77005, USA.

[2] Center for Theoretical Biological Physics, Rice University, Houston, TX 77005, USA.

[3] Department of Radiation Oncology, The University of Texas MD Anderson Cancer Center, Houston, TX 77030, USA.

[4] MD Anderson Morgan Welch Inflammatory Breast Cancer Research Program and Clinic, The University of Texas MD Anderson Cancer Center, Houston, TX 77030, USA.

[5] Department of Bioengineering, Rice University, Houston, TX 77005, USA.

[6] Department of Physics and Astronomy, Rice University, Houston, TX 77005, USA.

[*] Corresponding author. Email: mwdeem@rice.edu

# Abstract

**Background:** Clusters of circulating tumor cells (CTCs), despite being rare, may account for more than 95% of metastases. Cells in these clusters do not undergo a complete epithelial-to-mesenchymal transition (EMT) but retain some epithelial traits as compared to individually disseminating tumor cells. Determinants of single cell dissemination versus collective dissemination remain elusive. Inflammatory breast cancer (IBC), a highly aggressive breast cancer subtype that chiefly metastasizes via CTC clusters, is a promising model for studying mechanisms of collective tumor cell dissemination. Previous studies on breast cancer and adult acute myeloid leukemia, motivated by a theory that suggests physical systems with hierarchical organization tend to be more adaptable, have found that the expression of metastasis associated genes is more hierarchically organized in cases of successful metastases.

**Methods:** We used the cophenetic correlation coefficient (CCC) to quantify the hierarchical organization in the expression networks of two distinct gene sets, collective dissemination associated genes and IBC associated genes, in cancer cell lines and in tumor samples from breast cancer patients. Hypothesizing that a higher CCC for collective dissemination associated genes and for IBC associated genes would be associated with a more evident epithelial phenotype and with worse outcomes in breast cancer patients, we evaluated the correlation of CCC with different phenotypic groups.

**Results:** The CCC of both gene networks, the collective dissemination associated gene network and the IBC associated gene network, was higher in (a) epithelial cell lines as compared to mesenchymal cell lines and (b) tumor samples from IBC patients, as compared to samples from non-IBC breast cancer patients. A higher CCC of both networks was also correlated with a higher rate of metastatic relapse in breast cancer patients. Neither the levels of CDH1 gene expression, nor gene set enrichment analysis could provide similar insights.

**Conclusions:** These results suggest that retention of some epithelial traits in disseminating tumor cells as IBC progresses promotes successful breast cancer metastasis to distant organs. The CCC provides additional information regarding the organizational complexity of gene expression in comparison to differential gene expression analyses. We have shown that the CCC may be a useful metric for investigating the collective dissemination phenotype and a prognostic factor for IBC.

**Keywords:** Collective dissemination – Inflammatory Breast Cancer – EMT – Hierarchy – E/M hybrid – CCC

# Background

Metastasis is responsible for 90% of deaths from solid tumors [1]. It involves the escape of cancer cells from the site of the primary tumor, their entry into the circulatory system, and finally, colonization of and proliferation at a distant organ. However, this process is highly inefficient. Only an estimated 0.2% of the disseminated tumor cells are able to form a lesion at distant organ sites [2, 3]. A well-studied mechanism of metastasis is single cell dissemination where carcinoma cells acquire migratory and invasive traits via an epithelial-to-mesenchymal transition (EMT) [4]. These cells can then utilize blood or lymph circulation to reach distant organ sites, where they reacquire epithelial traits of cell-cell adhesion and apico-basal polarity via a mesenchymal-to-epithelial transition (MET) to establish metastases [4].

Recent studies have called into question the indispensability of a complete EMT and MET in metastasis [5] and have suggested collective migration of tumor cells via clusters of circulating tumor cells (CTCs) as an alternate mechanism of metastasis [6]. Clusters of tumor cells have been detected in the bloodstream of cancer patients even before the characterization of EMT as a driver of cancer metastasis [7, 8]. These clusters of tumor cells can efficiently seed secondary tumors, exhibiting up to 50 times the metastatic potential of individually migrating tumor cells [9]. Tumor cell clusters accounted for >90% of metastases in a mouse model of breast cancer [10]. Abundance of CTC clusters in the bloodstream has been associated with significantly poor prognosis in breast cancer and in small cell lung cancer [9, 11]. Multiple factors are believed to be responsible for the heightened metastatic potential of these CTC clusters. These include effective response to mechanical signals and chemical gradients by cells in CTC clusters as compared to migrating single tumor cells [12, 13], better evasion of the host immune system [14], and potential cooperation among heterogeneous cell types in CTC clusters [15]. Studies have shown that collectively invading tumor cells from the primary lesion often co-express epithelial and mesenchymal markers [16–18]. Thus, cells in CTC clusters tend to manifest a hybrid epithelial / mesenchymal (E/M hybrid) phenotype and to retain cell-cell adhesion characteristics [6].

Inflammatory Breast Cancer (IBC) is a highly aggressive breast cancer subtype that has been reported to predominantly metastasize via CTC clusters [19]. Characterized by breast erythema, edema, and *peau d'orange* presenting with or without a noticeable tumoral mass [20, 21], IBC involves tumoral infiltrate in the dermal lymphatics, and about 30% of IBC patients have distant metastases at the time of diagnosis as compared to only 5% of non-IBC type breast cancer patients [22]. Though only 2-4% of breast cancer cases each year in the United States are of the IBC type, IBC patients account for 10% of the annual breast cancer related mortalities. A hallmark of IBC is the presence of cohesive clusters of tumor cells in the local lymph nodes [23], and IBC patients have larger and a higher frequency of CTC clusters as compared to non-IBC breast cancer patients [24]. Abundance of CTC clusters has been shown to be associated with poor progression free survival in IBC patients [24]. Despite their great propensity to metastasize, tumor cells in the primary lesion and in metastatic lesions of IBC maintain a high expression E-cadherin, a hallmark of epithelial cells [23]. IBC thus presents an exciting model for the study of collective dissemination of tumor cells via CTC clusters and of the prognostic potential of these clusters of migrating tumor cells. The results we present later in this paper strengthen the argument for investigating IBC to elucidate the mechanisms underlying collective dissemination of tumor cells.

Here, we invoke concepts from theoretical models of evolution to investigate cluster-based dissemination of tumor cells and analogous IBC characteristics. Theoretical studies suggest that systems with a more hierarchical structure are more adaptable [25–27] due to their ability to efficiently span the space of possible states and are more robust to perturbations since a more hierarchical network structure has a buffering effect that hinders the propagation of local perturbations to a majority of nodes [27, 28]. Hierarchical organization, thus, emerges over time in physical systems evolving in a changing environment with a rugged fitness landscape exhibiting numerous peaks and valleys [26]. Given that tumor cells involved in metastasis and invasion progress through many different microenvironments [29–31], one can expect the expression of genes associated with a metastatic phenotype to be more hierarchically organized in instances of successful macrometastases as compared to instances with no metastasis.

Here, we quantify the hierarchical organization in the expression of two distinct sets of genes, one associated with collective dissemination of tumor cells and the other related to IBC, in cancer cell lines and in breast cancer patients. For this purpose, we use the use the cophenetic correlation coefficient (CCC) metric. A higher CCC indicates greater hierarchical organization in the expression of genes. The CCC was first used for comparing tree-like relationships represented by different dendrograms [32]. It has been used previously to quantify the difference in expression of metastasis associated genes in breast cancer patients with different clinical outcomes [33] and to quantify the difference in expression of genes predictive of clinical outcome in adult acute myeloid leukemia in patients belonging to different risk categories [34].

The first set of genes investigated here includes 87 genes reported to be associated with collective dissemination of tumor cells as CTC clusters: genes differentially expressed in CTC clusters as compared to individual CTCs [10]. The second gene set includes 78 genes reported to be differentially expressed in IBC patients in comparison to non-IBC breast cancer patients [35]. We observed that the CCC for both of these gene sets is higher in (a) epithelial cell lines as compared to mesenchymal cell lines, and (b) IBC patients as compared to non-IBC breast cancer patients. A higher CCC further correlated with worse disease progression in breast cancer patients. In light of these observations, we propose that the metastatic aggressiveness of IBC potentially derives from the hierarchical organization in the expression of collective dissemination associated genes in metastasizing tumor cells.

# Methods

**Genes associated with collective dissemination of tumor cell clusters**

Using multicolor lineage tracking, Cheung *et al.* showed that polyclonal seeding by disseminated clusters of tumor cells is the dominant mechanism for metastasis in a mouse model of breast cancer [10]. These clusters accounted for more than 90% of distant organ metastases in mice [10]. Circulating tumor cell clusters were observed to be enriched in expression of the epithelial protein keratin 14 (K14), and 87 genes with enriched or depleted expression in $K14^+$ cells as compared to $K14^-$ primary tumor cells were identified. Broadly, expression of adhesion complex associated genes was enriched and that of MHC Class II genes was depleted in $K14^+$ cells. We used this set of genes as a signature of the collective dissemination phenotype.

**Genes associated with the Inflammatory Breast Cancer (IBC) phenotype**

Van Laere *et al.* obtained tumor samples from patients with breast adenocarcinoma: 137 samples from IBC patients and 252 samples from patients with non-IBC type breast cancer (non-IBC) [35]. IBC patients were selected in accordance with the consensus diagnostic criteria described by Dawood *et al.* [20]. RNA from the tumor samples was hybridized onto Affymetrix GeneChips (HGU133-series) to obtain the corresponding mRNA expression profiles. Linear regression models were employed to identify a set of 78 IBC specific genes which were differentially expressed in IBC tumor samples as compared to non-IBC tumor samples, independent of the molecular subtype of the tumor [35]. We used this set of genes as a signature of the IBC phenotype in breast cancer patients.

**Gene expression data from different cell lines**

We used two different datasets of gene expression in cell lines, each cell line classified as epithelial (E), mesenchymal(M), or epithelial / mesenchymal hybrid (E/M hybrid). The first dataset is from the study by Grosse-Wilde et al. [36]. A total of 24 clones established from HMLER cell lines (normal human mammary epithelial cells immortalized and transformed with hTERT and the oncogenes SV40LT and RAS [37]) were sorted into 13 CD24+/CD44- E clones and 11 CD24-/CD44+ M clones. The E clones and M clones displayed cobble-stone like morphology and dispersed, fibroblast morphology, respectively.

The second dataset includes gene expression from the National Cancer Institute (NCI) 60 anticancer drug screen (NCI-60), which includes panels of cell lines representing 9 distinct types of cancer: leukemia, colon, lung, central nervous system, renal, melanoma, ovarian, breast, and prostate [38]. The 60 cell lines have been classified into epithelial (E) (n = 11), mesenchymal (M) (n = 36), and epithelial / mesenchymal hybrid (E/M) (n = 11) categories on the basis of protein levels of E-cadherin and Vimentin [39]

**Gene expression data from tumor samples from IBC and non-IBC breast cancer patients**

We used three different datasets of gene expression in tumor samples obtained from breast cancer patients. Each patient in the three datasets was diagnosed with either inflammatory breast cancer (IBC) or non-IBC type breast cancer (non-IBC). Iwamoto *et al.* collected tumor biopsies prospectively from 82 patients with locally advanced disease. A clinical diagnosis of IBC was made in 25 of these patients [40]. Boersma *et al.* examined primary breast tumor samples from 50 patients, 15 of whom were diagnosed with IBC on the basis of the pathology and medical reports [41]. Finally, Woodward *et al.* obtained tissue samples from core biopsies of breast tissue in 40 breast cancer patients, 20 IBC and 20 non-IBC [21].

In Iwamoto *et al.* and Woodward *et al.*, IBC diagnosis was made in patients with clinical presentation of breast erythema and edema over more than one-third of the breast. In Boersma *et al.*, 9 IBC patients presented with erythema and edema, while 6 IBC patients exhibited pathology indicating dermal lymphatic invasion and tumor emboli.

**Definition of gene network for different phenotypic groups**

For each phenotypic group, e.g. NCI60 cell lines labeled as epithelial (E) or patients in the Iwamoto *et al.* [40] dataset diagnosed with IBC, and gene set, e.g. the gene set associated with IBC or the set of collective dissemination associated genes, we defined a network with the genes as nodes and weighted edges between these nodes. The weight of the edge between gene $i$ and gene $j$ in the phenotypic group $G$ was defined as

$$l_{ij}^G = \left| \sum_{k \in G} \frac{(e_i^k - \mu_i^G)(e_j^k - \mu_j^G)}{\sigma_i^G \sigma_j^G} \right| \quad (1)$$

Here, $e_m^k$ is the expression of gene $m$ in the sample $k$ (patient / cell line), $\mu_m^G$ and $\sigma_m^G$ are the mean and standard deviation of the expression of gene $m$ in the phenotypic group $G$ respectively, and the summation is over all the patients or cell lines belonging to the group $G$.

We constructed such networks for the epithelial and mesenchymal cell lines in the Grosse-Wilde *et al.* [36] dataset and for the epithelial, mesenchymal, and epithelial / mesenchymal hybrid cell lines in the NCI60 dataset. Such networks were also constructed for IBC and non-IBC patients in the three breast cancer datasets, Iwamoto *et al.* [40], Woodward *et al.* [21], and Boersma *et al.* [41], using each of the two gene sets described previously, genes associated with collective dissemination of tumor cell clusters and genes associated with the IBC phenotype.

**Calculation of the Cophenetic Correlation Coefficient**

To quantify the hierarchy in the expression of the two sets of genes in different groups of patients and cell lines, we used a metric called the cophenetic correlation coefficient (CCC) [32]. The CCC is a measure of how well a hierarchical clustering of nodes in a network reproduces the distances between nodes in the original network. Intuitively, the CCC is a measure of how tree-like a network is. Since a tree topology is a prototypical hierarchical structure, a measure of the tree-like characteristic of a network allows us to aptly quantify the underlying hierarchy in the structure of a network.

For calculating the CCC of a given network, we defined the distance between nodes $i$ and $j$, $d_{ij}$, as the Euclidean commute time distance (ECTD) between the nodes $i$ and $j$ [42]. The ECTD between nodes $i$ and $j$ depends not only on the weight of the edge between nodes $i$ and $j$, but also on the number of different possible paths between the two nodes. The ECTD decreases as the number of possible paths between the two nodes increases, and increases if any path between the two nodes becomes longer. This makes the ECTD suitable for clustering tasks. For a network with $N$ nodes, we generated a $N \times N$ matrix $D$ such that $D_{ij}$ is the ECTD between nodes $i$ and $j$ [43]. The matrix $D$ is then used as an input to the average linkage hierarchical clustering algorithm [44] which generates a tree topology ($T$), i.e. a dendrogram, that best approximates the distances between the nodes of the network given by the matrix $D$. We then calculated the CCC as the correlation between the original pairwise distances and the corresponding distances in the tree topology:

$$CCC = \frac{\sum_{i<j}(D_{ij} - d)(T_{ij} - t)}{\sqrt{\sum_{i<j}(D_{ij} - d)^2 \sum_{i<j}(T_{ij} - t)^2}} \quad (2)$$

Here, $d = <D_{ij}>$ is the mean of the original pairwise distances and $t = <T_{ij}>$ is the mean of the pairwise distance in the tree topology. If the original network is hierarchical, the distances between nodes in the tree topology obtained via hierarchical clustering ($T$) will be highly correlated with the distances between nodes in the original network ($D$). Hence, the CCC will be high. However, if the original network lacks any hierarchical organization, this correlation will be weak, and the CCC will be low.

The CCC calculated for a network was normalized with respect to the CCC of random networks with the same set of nodes but re-distributed edge weights. For this, we generated 10 such random networks by shuffling entries in the matrix $D$ and then calculated the average of the CCCs of these random networks (CCC_rand). The normalized CCC was then defined as

$$CCC_{norm} = \frac{CCC - CCC_{rand}}{1 - CCC_{rand}} \quad (3)$$

Finally, to obtain the error in the estimate of CCC, we used the bootstrap method [45]. The method assumes that the distribution of gene expression in a patient or cell line group is the empirical distribution function of the observed expression of samples within the group. For a patient or cell line group with size $p$, we drew $p$ samples from the group with replacement and calculated CCC_norm for the sampled group. This sampling process was repeated 100 times to obtain 100 CCC_norm values. The standard error in the estimate of the CCC_norm for the group was then given as the sample standard deviation of the 100 sampled CCC_norm values.

# Results

**Higher CCC for the collective dissemination associated gene network in epithelial cell lines and in IBC patients.**

We constructed networks with genes associated with collective dissemination of tumor cell clusters [10], hereafter referred to as 'collective dissemination associated' genes, as nodes and weights of the edges between a pair of nodes defined according to equation (1). Such networks were constructed for the E and M cell lines from the gene expression data from Grosse-Wilde *et al.* [36] and for the cell lines in the NCI-60 anti-cancer drug screen [38] that have been categorized into E, M, and E/M hybrid classes [39]. The normalized CCC for these networks was calculated using the method described above, and the results are shown in fig. 1. E cell lines exhibited a significantly higher CCC as compared to M cell lines (p-value < 0.05) for the collective dissemination associated gene network in the dataset from Grosse-Wilde *et al.* [36], fig. 1 (A). In the NCI60 dataset, the CCC of the collective dissemination associated gene network was higher for E cell lines as compared to the pooled M and E/M hybrid cell lines, fig. 1(B). The bootstrap distribution of normalized CCC values for E cell lines was distinct from the distribution for M cell lines in the dataset from Grosse-Wilde *et al.* [36] and from the distribution for pooled M and E/M hybrid cell lines in the NCI-60 dataset (Kolmogorov–Smirnov test, p-value < $10^{-5}$).

We constructed similar networks for IBC and non-IBC patients using Affymetrix U133A profiles obtained by Iwamoto *et al.* [40]. Normalized CCC values for patients in the two groups are shown in fig. 2 (A). IBC patients exhibited a higher CCC for the network associated with collective dissemination of tumor cell clusters as compared to non-IBC breast cancer patients. The difference between the two groups in the dataset was significant (p-value < 0.05). Further, bootstrap distributions for the normalized CCC values for the two are groups were statistically distinct with p-value < $10^{-5}$ for the Kolmogorov–Smirnov test). However, we did not observe a significant trend for the breast cancer samples characterized by Boersma *et al.* [41] and for the samples characterized by Woodward *et al.* [21], fig. 2 (B) and (C).

**Higher CCC for the IBC associated gene network in epithelial cell lines and in IBC patients.**

We constructed networks with genes differentially expressed in tumor samples obtained from IBC patients as compared to tumor samples from non-IBC breast cancer patients, hereafter referred to as 'IBC associated' genes, as nodes. Weights of edges between pairs of nodes were defined using equation (1). Such networks were constructed for the E and M cell lines in the dataset from Gross-Wilde *et al.* [36] and for the E and pooled M + E/M hybrid cell lines in the NCI-60 dataset. Normalized CCC values for these groups of cell lines calculated using the method described above are shown in fig. 3. E cell lines displayed a higher CCC for the IBC associated gene network as compared to other cell lines in both datasets (p-value < 0.05 in each case). The bootstrap distributions of normalized CCC values for the two groups of cell lines were statistically distinct for both datasets ($p < 10^{-5}$ for the Kolmogorov–Smirnov test in each case).

Using Affymetrix U133A profiles from Iwamoto *et al.* [40], we constructed similar networks with IBC associated genes as nodes for both IBC and non-IBC breast cancer patients. Normalized CCC values for the two breast cancer patient groups are shown in fig. 4. The IBC group exhibited a significantly higher CCC for the IBC associated gene network as compared to the non-IBC patients group (p-value < 0.05). Bootstrap distributions for the two groups were again statistically distinct (p-value $< 10^{-5}$ for the Kolmogorov–Smirnov test), fig. 4 (A). A similar trend in the CCC values for IBC and non-IBC patient groups was observed for breast cancer patients in the two other independent breast cancer datasets, Woodward *et al.* [21] and Boersma *et al.* [41], fig. 4 (B) and (C).

Saunders and McClay used a well-understood gene regulatory network in the sea urchin embryo to identify transcription factors that control cell changes during EMT by perturbing individual transcription factors [46]. They further determined 30 human transcription factors homologous to those identified in sea urchins. We calculated the CCC of a network with these transcription factors, hereafter referred to as 'canonical drivers of EMT', as nodes for the IBC and non-IBC samples from each of the three breast cancer datasets, Iwamoto *et al.* [40], Boersma *et al.* [41], and Woodward *et al.* [21]. The weights of edges between different transcription factors were defined using equation (1). We observed that the IBC patient group exhibited a lower CCC for the network composed of canonical EMT drivers as compared to the non-IBC patient group in data from each of the three studies, fig. 5.

**Higher CCC for the two networks correlates with a higher rate of metastasis**

We constructed networks with the two sets of genes, collective dissemination associated and IBC associated, as nodes for breast cancer patients who exhibited metastatic relapse within 5 years post-treatment, Wang *et al.* [47]. These patients were classified into two groups, those with metastatic relapse within 30 months and those with metastasis between 30 to 60 months post-treatment. Edge weights were defined, once again, using equation (1). For both collective dissemination of tumor cells associated and IBC associated gene sets, the CCC was significantly higher (p < 0.05) for the patient group with early metastatic relapse of breast cancer, i.e. relapse within 30 months of treatment, as compared to patients with relatively late relapse, i.e. metastatic relapse after 30 months post-treatment, fig. 6 (A) and (B). The same trend was observed upon considering only estrogen-receptor-positive patients, fig. S1. There were too few samples from estrogen-receptor-negative patients for similar analysis. A similar trend was observed for small cell lung cancer (SCLC), another highly aggressive cancer subtype, which has also been reported to metastasize via cluster-based dissemination of tumor cells [48]. Patients with fewer than 10 months of disease free survival post-treatment exhibited a higher CCC for both collective dissemination associated and IBC associated gene sets as compared to patients with greater than 10 months of disease free survival post-treatment as computed from the data in the study by Rousseaux *et al.* [49], fig. 6 (C) and (D). We also compared CCCs of both collective dissemination associated and IBC associated gene networks in samples from breast cancer metastases to different organs and observed a higher CCC for metastases to skin and liver as compared to metastases to lymph nodes in the data from Kimbung *et al.* [50], fig. 7.

We further explored whether the CCC for the collective dissemination associated gene network and the IBC associated gene network were different in breast cancer patients with metastatic relapse within 5 years post-treatment and those with no metastasis during this follow-up period as computed from the data in the study by Wang *et al.* [47]. Intriguingly, we observed that the CCC of both networks was significantly higher ($p < 0.05$ in each case) for patients with no metastasis during the 5-year follow up period as compared to patients with metastatic relapse during the follow up, fig. 8 (A) and (B). A similar trend was observed for breast tumor samples from the cancer genome atlas (TCGA) for patients who exhibited relapse during the follow up period and those who did not [51], fig. 8 (C) and (D). Given that healthy breast cells are inherently epithelial, a higher CCC for the patient group with no metastatic relapse during the follow up period may be a consequence of the tumor being at initial stages of progression towards a metastatic phenotype at the time of diagnosis and sample collection in these patient groups. However, upon grouping the breast cancer patients by their estrogen-receptor status, no consistent trend was observed between patients with no relapse during the 5-year follow-up period and patients with metastatic relapse within 5 years post-treatment for both gene sets, fig. S2. These results indicate that the collective dissemination pathway in breast cancer patients with differing receptor statuses warrants further study.

**The CCC provides additional information regarding the underlying complexity of collective gene expression**

We investigated if the insights described above can be obtained from a straightforward analysis of gene expression levels. To determine how the CCCs of different gene networks correlate with the expression of these genes in different phenotypic groups, we carried out gene set enrichment analysis (GSEA) for different sets of genes on epithelial and mesenchymal cell lines from the study by Grosse-Wilde *et al.* [36] and on the tumor samples from IBC patients and non-IBC breast cancer patients from the study by Iwamoto *et al.* [40]. Using the GSEA software provided by the Broad Institute [52], we tested for enrichment in the expression of collective dissemination associated genes, IBC associated genes, and of the canonical drivers of EMT in different phenotypic groups, i.e. epithelial versus mesenchymal cell lines in the data from Grosse-Wilde *et al.* [36] and IBC versus non-IBC patients in the data from Iwamoto *et al.* [40]. The results are shown in fig. 9 (A-F). The expression of collective dissemination associated genes is significantly enriched in epithelial cell lines as compared to mesenchymal cell lines (p-value < 0.001) while IBC associated genes and canonical EMT drivers do not show significant enrichment when compared across these two phenotypic groups. On the other hand, expression of IBC associated genes is significantly enriched in tumor samples from IBC patients (p-value =

0.035) while the collective dissemination associated genes and canonical EMT drivers do not show significant enrichment on comparing IBC samples with non-IBC breast tumor samples.

Previous studies have suggested a strong association between expression of E-cadherin protein in tumor cells and IBC [53, 54]. We compared the levels of CDH1 (E-cadherin) gene expression in tumor samples from IBC and non-IBC patients. There was no significant difference in expression levels of CDH1 gene between the two groups in any of the 3 breast cancer patients datasets, Iwamoto *et al.* [40], Boersma *et al.* [41], and Woodward *et al.* [21], fig. 9 (G-I).

Together, these results indicate the CCC need not correlate with differential gene expression analysis. In fact, the CCC of a set of genes for two samples with a k-fold change in the expression of all genes in the set will be the same. On the other hand, a k-fold change in the expression levels of all genes in a set as compared to the background will immensely change the gene set enrichment score for that set of genes. The CCC can thus provide insights in addition to those that may be obtained from a direct analysis of gene expression data by using GSEA.

# Discussion

Cancer metastasis via migrating clusters of circulating tumor cells has emerged as a critical mechanism of seeding secondary tumors in recent studies [7–10]. Although rare in comparison with individually disseminated cancer cells, CTC clusters can efficiently seed secondary tumors at distant organ sites [9, 10], and their presence in the bloodstream of cancer patients has been shown to be associated with poor disease prognosis, i.e. worse overall survival and worse disease-free survival [9]. Understanding the molecular mechanisms underlying collective dissemination of tumor cells is, therefore, important for predicting metastasis, which remains the principal cause of cancer associated mortalities. Determinants of single cell versus collective dissemination of tumor cells, however, remain elusive. Here, we have analyzed the topology of the network of genes implicated in collective dissemination of tumor cell clusters. We also investigated the topology of the network of genes reported to be differentially expressed in patients with inflammatory breast cancer (IBC), a highly aggressive type of breast cancer characterized by lymphatic emboli composed of clusters of tumor cells. Taken together, our analysis suggests that maintenance of the epithelial phenotype in cancer cells disseminating from the primary tumor contributes towards metastasis via collective migration of tumor cells as CTC clusters.

Results suggest that expression of genes differentially expressed in tumor cells migrating as clusters as compared to individually migrating tumor cells [10] exhibits a more hierarchical organization in epithelial cell lines as compared to mesenchymal cell lines in both, immortalized breast cell lines [36] and in the NCI-60 panel of cancer cell lines [38, 39]. Retention of some epithelial characteristics by cancer cells disseminating from the primary tumor has been reported to contribute towards collective invasion by tumor cells as CTC clusters [10, 55, 56]. A more hierarchical organization in the expression of these genes may contribute towards a more robust epithelial phenotype in these cell lines [25–28]. Higher hierarchical organization in the expression of these genes is also observed in tumor samples from IBC patients as compared to tumor samples from non-IBC breast cancer patients. This difference may contribute towards the strengthened presentation of epithelial characteristics, cell-cell adhesion and inter-cellular communication, in tumor cells from IBC patients that fosters the collective migration of these cells from the primary breast lesion [55]. Further, hierarchical expression of collective dissemination associated genes is of diagnostic relevance in IBC, thereby strengthening the case for IBC as a model system for the study of collective dissemination of tumor cells [19] and indicating the usefulness of mechanistic studies of tumor cell dissemination in determining the principles underlying IBC.

Next, we investigated the hierarchical organization in the expression of genes previously reported to be differentially expressed in tumor samples from IBC patients as compared to non-IBC patients [35]. The expression of these genes was more hierarchically organized in IBC samples as compared to non-IBC samples across multiple independent datasets. Further, epithelial cell lines exhibited a more hierarchical expression of these genes as compared to mesenchymal cell lines in both immortalized breast cell lines [36] and in the NCI-60 cell line panel composed of 9 different tumor types [38, 39]. Thus, both collective dissemination associated and IBC-associated genes exhibited a similar trend of higher CCC in immortalized breast cell lines or cancer cell lines as well as in tumor samples from breast cancer patients, adding to the existing evidence on collective dissemination via tumor emboli as the predominant mode of IBC metastasis and consequent aggressiveness. Intriguingly, the expression of canonical EMT inducing transcription factors [46] was more hierarchically organized in non-IBC breast cancer samples as compared to IBC samples. Taken together, these results reinforce the notion that a complete EMT is not involved in IBC metastasis. Rather, it is the collective migration of tumor cells that are able to retain some epithelial characteristics that contributes towards the metastatic aggressiveness of IBC.

Both collective dissemination associated and IBC associated gene sets exhibited a higher CCC in breast cancer patients with faster post-treatment metastatic relapse as compared to patients with slower post-treatment relapse [47]. A similar trend was observed in our calculations of the CCC for patients with SCLC [49], another metastatically aggressive cancer reported to metastasize via clusters of tumor cells [48]. These results indicate that a more hierarchical organization in the expression of genes involved in the collective dissemination of tumor cells may contribute towards a more aggressive behavior in metastatically aggressive tumors such as IBC and SCLC, which predominantly metastasize via clusters of circulating tumor cells. A mechanism based investigation of the cross-talk between collective dissemination associated and IBC associated genes may, therefore, be a promising next step. Further, samples from breast cancer metastases to lymph nodes exhibited a lower CCC as compared to breast cancer metastases to skin and liver for collective dissemination associated and IBC associated gene sets

[50]. While metastasis of tumor cells to distant organs is a complex, multi-step, and highly inefficient process, migration of tumor cells from the primary tumor to local lymph nodes is a more facile process and can be brought about by passive flow of the lymph. Correlation of the CCC for both gene sets, collective dissemination associated and IBC associated, with a higher rate of and propensity for metastasis to distant organs clearly speaks of the survival advantage afforded to migrating tumor cells by collective dissemination as clusters of CTCs. These advantages include enhanced ability to resist anoikis (cell death upon detachment from the substrate), evasion from immune system recognition, potential polyclonality, and enhanced ability to seed secondary tumors [57].

A commonly used approach to determine if an *a priori* defined set of genes is associated with phenotypic differences between two groups is gene set enrichment analysis (GSEA) [58, 59]. This method involves finding if the given set of genes is over-represented among genes that are differentially expressed in the two phenotypic groups. To determine if insights similar to those described above can be obtained via GSEA for the collective dissemination associated gene set and for the IBC associated gene set, we used the GSEA software provided by the Broad Institute [52] to calculate enrichment scores for the two gene sets in the data from Grosse-Wilde *et al.* [36], i.e. epithelial versus mesenchymal cell lines, and in the data from Iwamoto *et al.* [40], i.e. IBC versus non-IBC patients. While we consistently obtained a higher CCC for collective dissemination associated and IBC associated gene sets in both epithelial cell lines and tumor samples from IBC patients, the expression of these gene sets was not always enriched in epithelial versus mesenchymal cell lines or IBC versus non-IBC analysis. These results, thus, indicate that the CCC of a gene network can be a robust metric of functional significance of a set of genes in different phenotypic groups, independent of the enrichment score calculated for the given gene set. The CCC provides a prognostic measure based on the collective expression of genes in cells exhibiting different phenotypes beyond that provided by GSEA.

The classical view of cancer is that it involves de-differentiation of host cell pathways [60]. However, structure in the pathways involving genes that promote cancer progression may be selected for as the disease advances. We previously showed that the expression of adult acute myeloid leukemia associated genes is more hierarchically organized in samples from patients in whom the disease relapsed during the follow up period as compared to patients that underwent complete remission upon treatment [34]. Similarly, for breast cancer metastasis associated genes, hierarchical organization was higher in patients who developed distant metastases as compared to patients who did not [33]. Here, we propose that due to the role of maintenance of the epithelial phenotype in collective dissemination of tumor cells and the subsequent metastatic efficiency of CTC clusters, a hierarchical organization in the expression of these genes may be selected for in metastatically aggressive cancers like IBC. A measure of hierarchical organization, here the CCC, can thus be a useful biomarker in cancer prognosis, particularly in the case of IBC.

# Conclusions

We have shown that a set of genes previously reported to be associated with the collective dissemination of tumor cell clusters [10] is more hierarchically expressed in epithelial cell lines as compared to mesenchymal cell lines, thereby indicating a role for epithelial characteristics in the collective migration of tumor cells as clusters of circulating tumor cells. We further showed that IBC, an aggressive breast cancer subtype that metastasizes primarily via CTC clusters, exhibits a more hierarchical organization in the expression of these collective dissemination associated genes as compared to non-IBC type breast cancer. Along similar lines, we showed that for genes differentially expressed in IBC as compared to non-IBC type breast cancer, the expression is more hierarchical in tumor samples from IBC patients and in phenotypically epithelial cell lines, suggesting a role for the epithelial phenotype in the metastatically aggressive nature of IBC. Taken together, our work indicates that maintenance of the epithelial phenotype in disseminating tumor cells during disease progression plays a key role in successful metastasis of cancer to distant organs, and that IBC can be a suitable model system for studying mechanisms of collective migration of tumor cells and CTC clusters. Further, we have introduced the CCC as a quantitative metric for analyzing the collective migration of circulating tumor cell clusters, which can be useful in cancer prognosis, particularly in the case of IBC.

**Abbreviations**

EMT: epithelial-to-mesenchymal transition; CTC: circulating tumor cell; IBC: inflammatory breast cancer; CCC: cophenetic correlation coefficient; K14: keratin 14; ECTD: Euclidean commute time distance; SCLC: small cell lung cancer; GSEA: gene set enrichment analysis

**Ethical approval and consent to participate**

Not applicable.

**Consent for publication**

Not applicable.

**Availability of data and material**

All gene expression datasets analyzed during this study are publicly available on the Gene Expression Omnibus (GEO) database (https://www.ncbi.nlm.nih.gov/geo/). See text for relevant papers. Gene lists analyzed in the study are included in the corresponding articles (Cheung *et al.* [10] and Van Laere *et al.* [35]).

**Competing interests**

The authors declare that they have no competing interests.

**Funding**

This work was supported by the Center for Theoretical Biological Physics and funded by the National Science Foundation (PHY-1427654). MKJ has a training fellowship from the Gulf Coast Consortia on the Computational Cancer Biology Training Program (CPRIT grant no. RP170593).

**Authors' contributions**

ST designed the study, carried out the analysis, and wrote the paper. MKJ, WAW, HL, and MWD designed the study and wrote the paper.

**Acknowledgements**

Not applicable.

# Figures

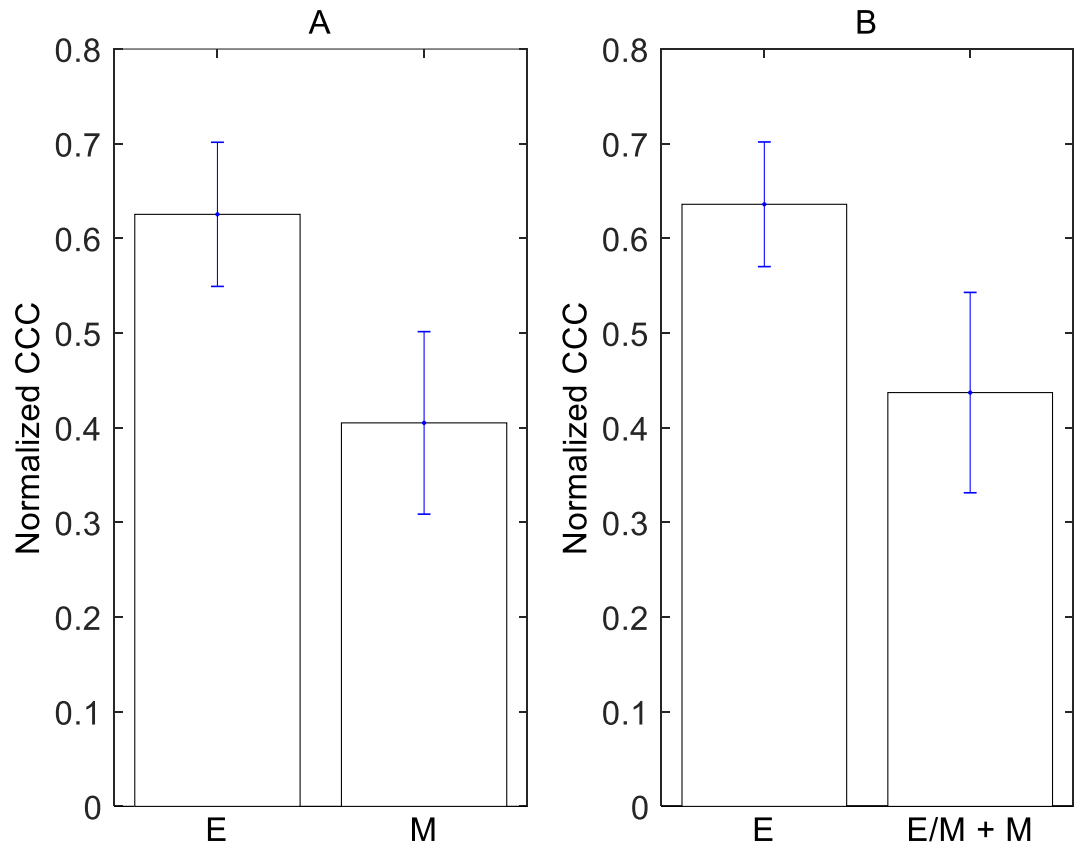

Figure 1 Normalized CCC for the collective dissemination associated gene network for cell lines from two different datasets: (A) epithelial (E) (n = 13) and mesenchymal (M) (n = 11) cell lines from the study by Grosse-Wilde *et al.* [36] and (B) epithelial (E) (n = 11) and epithelial / mesenchymal hybrid (E/M) + mesenchymal (M) (n = 47) cell lines from the NCI-60 dataset [38, 39]. In both datasets, the epithelial cell lines exhibit a higher CCC for collective dissemination associated gene network (p-value < 0.05). Error bars indicate the standard error in the estimate of $CCC_{norm}$ calculated using the bootstrap method.

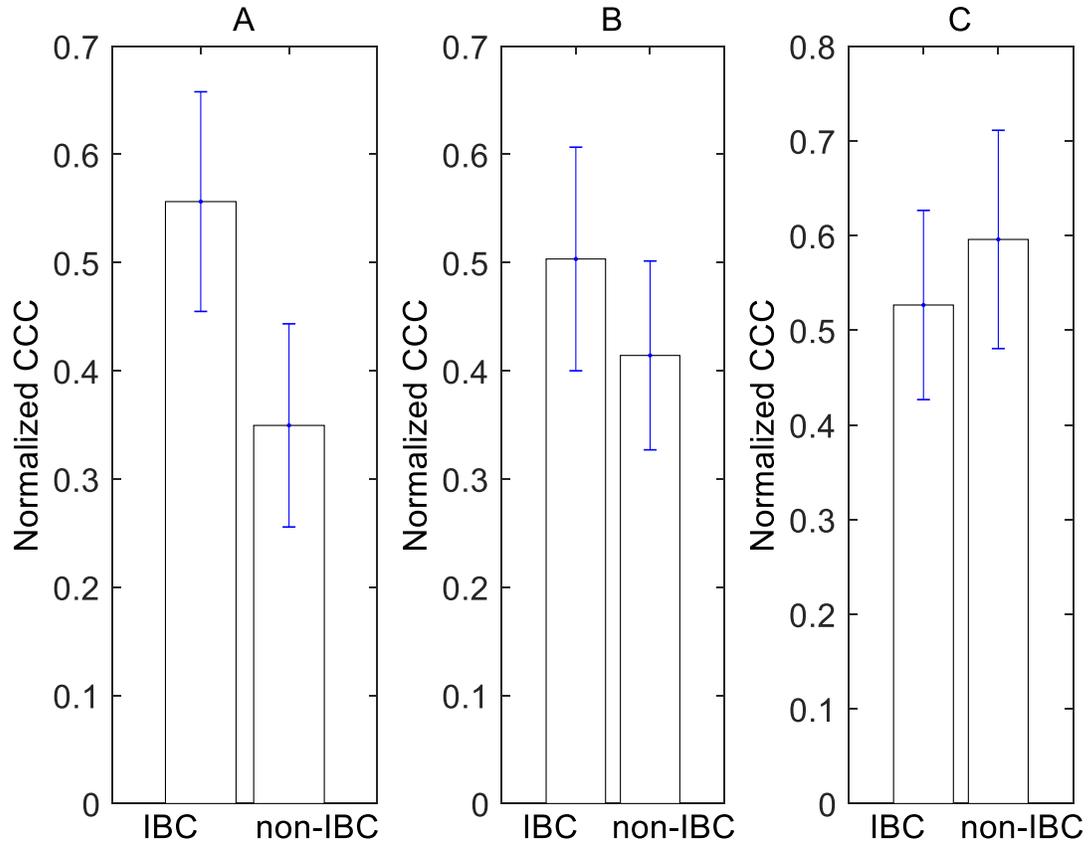

Figure 2 Normalized CCC of the collective dissemination associated gene network for tumor samples from IBC patients and non-IBC breast cancer patients for data from studies by (A) Iwamoto *et al.* [40] (n = 25 for the IBC group, n = 57 for the non-IBC group), (B) Boersma *et al.* [41] (n = 13 for the IBC group, n = 35 for the non-IBC group), and (C) Woodward *et al.* [21] (n = 20 for the IBC group, n = 20 for the non-IBC group). Error bars indicate the standard error in the estimate of $CCC_{norm}$ calculated using the bootstrap method.

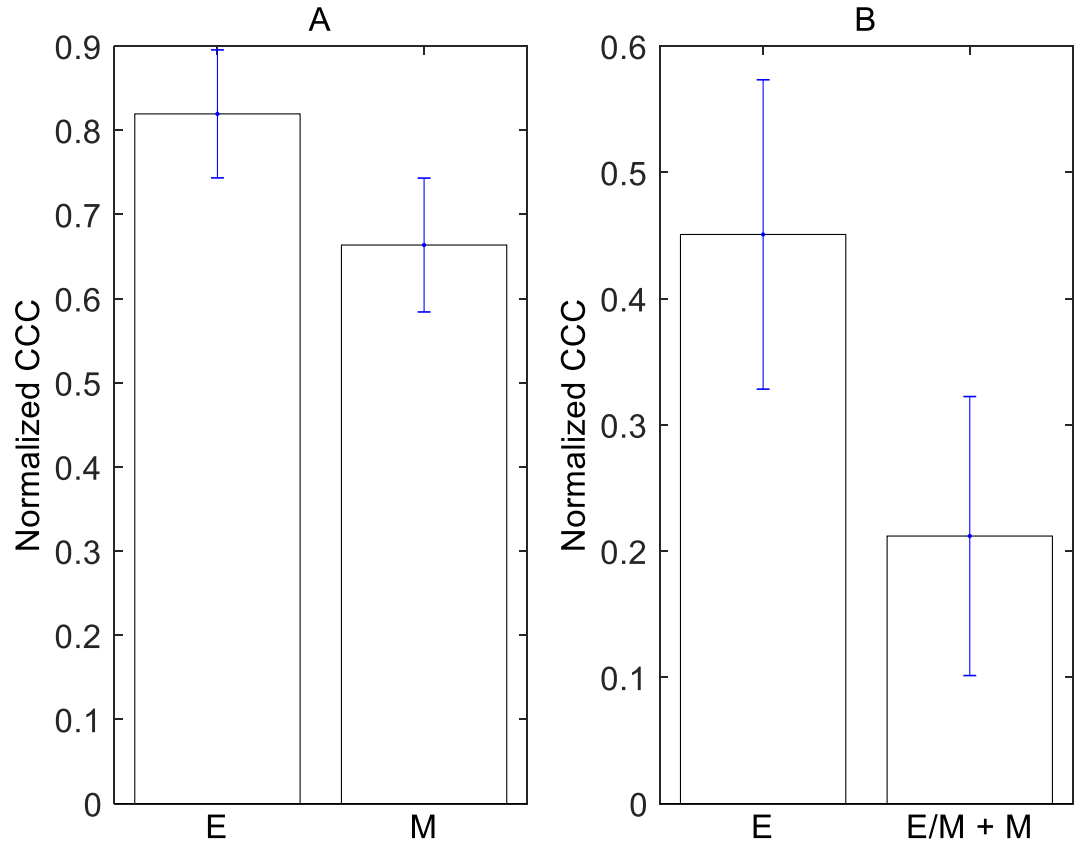

Figure 3 Normalized CCC for the IBC associated gene network for cell lines from two different datasets: (A) epithelial (E) (n = 13) and mesenchymal (M) (n = 11) cell lines from the study by Grosse-Wilde *et al.* [36] and (B) epithelial (E) (n = 11) and epithelial / mesenchymal hybrid (E/M) + mesenchymal (M) (n = 47) cell lines from the NCI-60 dataset [38, 39]. In both datasets, the epithelial cell lines exhibit a higher CCC for the IBC associated gene network (p-value < 0.05). Error bars indicate the standard error in the estimate of $CCC_{norm}$ calculated using the bootstrap method.

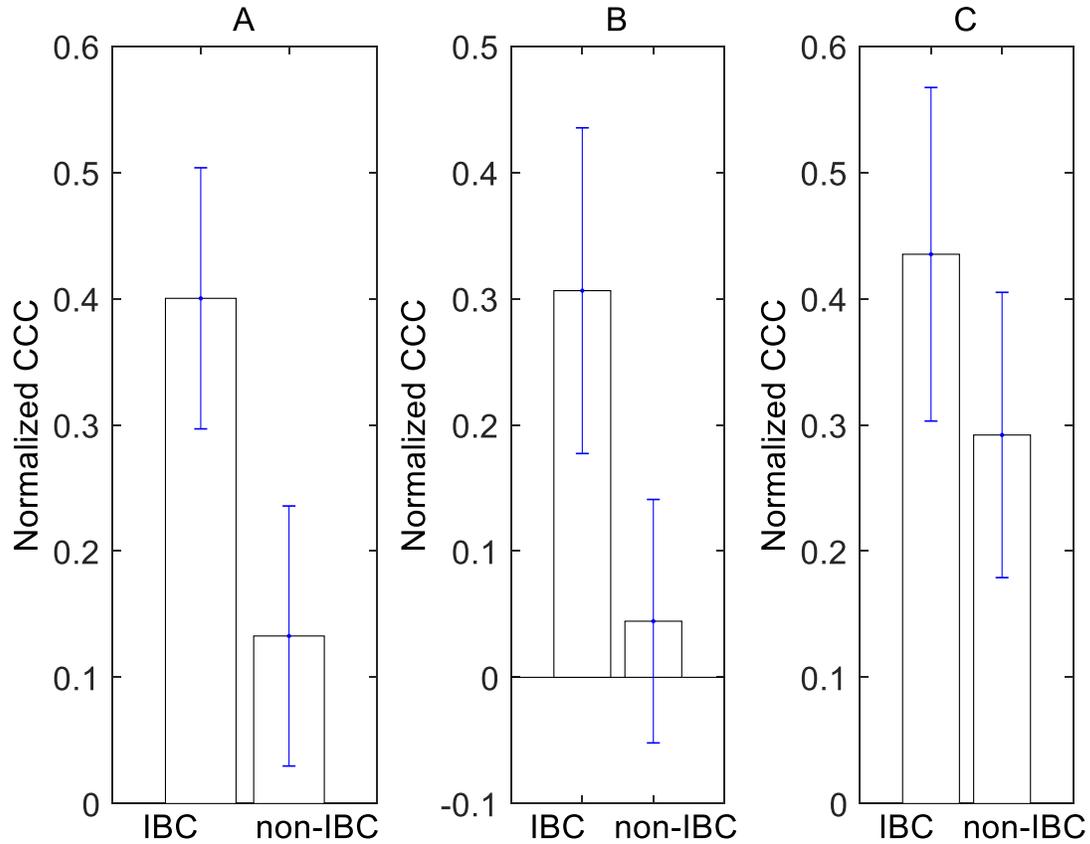

Figure 4 Normalized CCC of the IBC associated gene network for tumor samples from IBC patients and non-IBC breast cancer patients for date from studies by (A) Iwamoto *et al.* [40] (n = 25 for the IBC group, n = 57 for the non-IBC group), (B) Boersma *et al.* [41] (n = 13 for the IBC group, n = 35 for the non-IBC group), and (C) Woodward *et al.* [21] (n = 20 for the IBC group, n = 20 for the non-IBC group). Error bars indicate the standard error in the estimate of $CCC_{norm}$ calculated using the bootstrap method. Tumor samples from IBC patients exhibit a higher CCC for the IBC associated gene network as compared to tumor samples from non-IBC breast cancer patients (p-value < 0.05 for (A) and (B)).

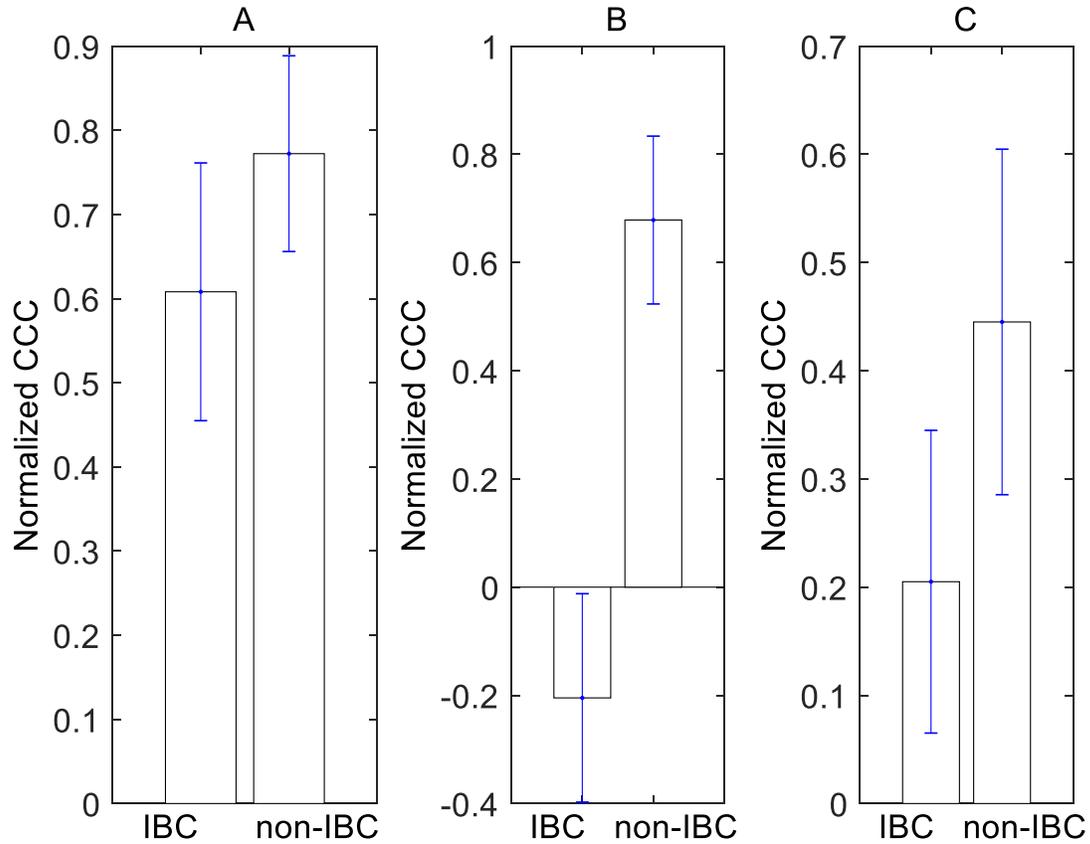

Figure 5 Normalized CCC for the expression of canonical EMT driving transcription factors in IBC and non-IBC breast cancer patient groups from studies by (A) Iwamoto *et al.* [40] (n = 25 for the IBC group, n = 57 for the non-IBC group), (B) Boersma *et al.* [41] (n = 13 for the IBC group, n = 35 for the non-IBC group), and (C) Woodward *et al.* [21] (n = 20 for the IBC group, n = 20 for the non-IBC group). Error bars indicate the standard error in the estimate of $CCC_{norm}$ calculated using the bootstrap method.

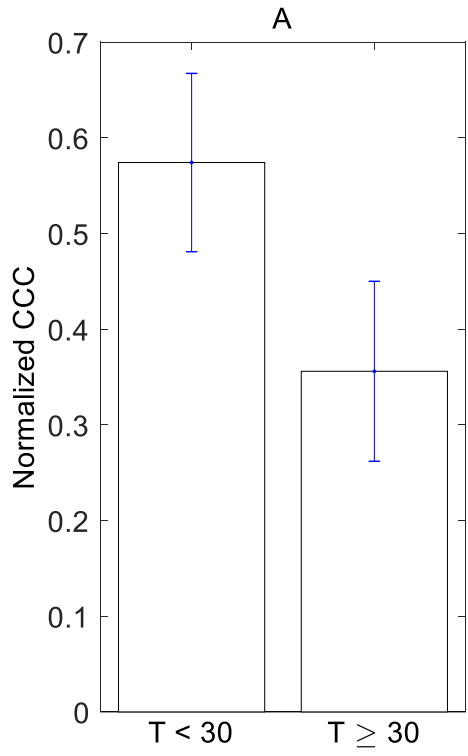
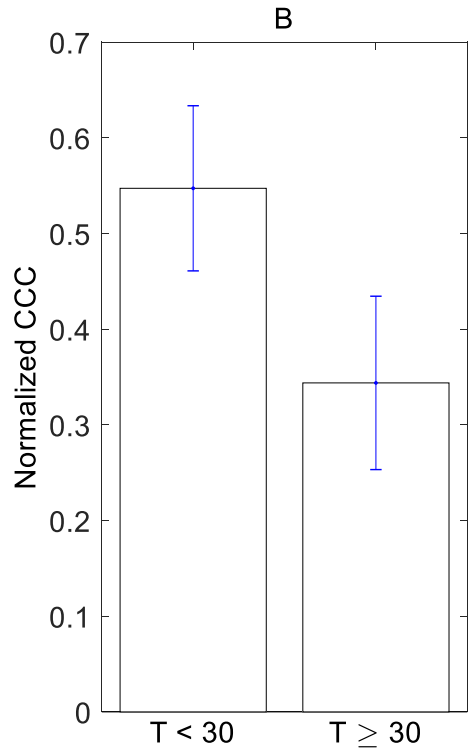
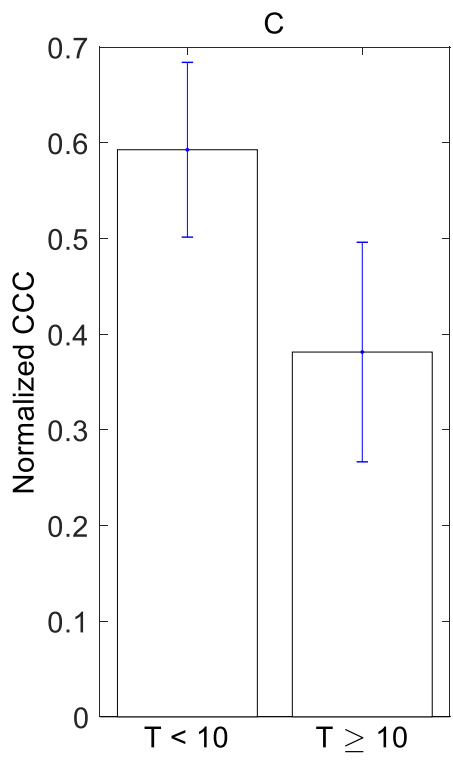
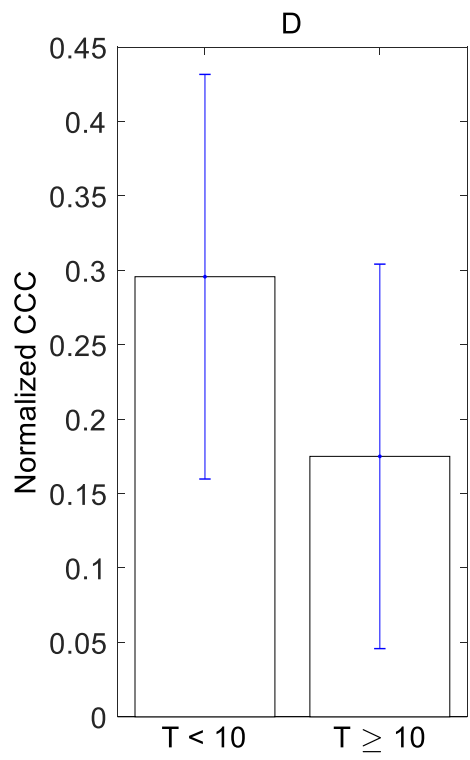

Figure 6 (A) Normalized CCC of the collective dissemination associated gene network for breast cancer patients with metastatic relapse within a 30-month period post-treatment (T < 30; n = 56) and with metastatic relapse between 30 and 60 months post-treatment (T ≥ 30; n = 51). Data from the study by Wang *et al.* [47]. (B) Normalized CCC of the IBC associated gene network for same groups of breast cancer patients as in (A). (C) Normalized CCC of the collective dissemination associated gene network for small cell lung cancer (SCLC) patients with less than 10 months of disease free survival post-treatment (T < 10; n = 11) and patients with longer than 10 months of disease free survival post-treatment but death during the follow up period (T ≥ 10; n = 10). Data from the study by Rousseaux *et al.* [49]. (D) Normalized CCC of the IBC associated network for the same SCLC patient groups as in (C). Higher CCCs of the collective dissemination associated and IBC associated gene networks correlate with a higher rate of metastasis, both in breast cancer and small cell lung cancer. Error bars indicate the standard error in the estimate of $CCC_{norm}$ calculated using the bootstrap method.

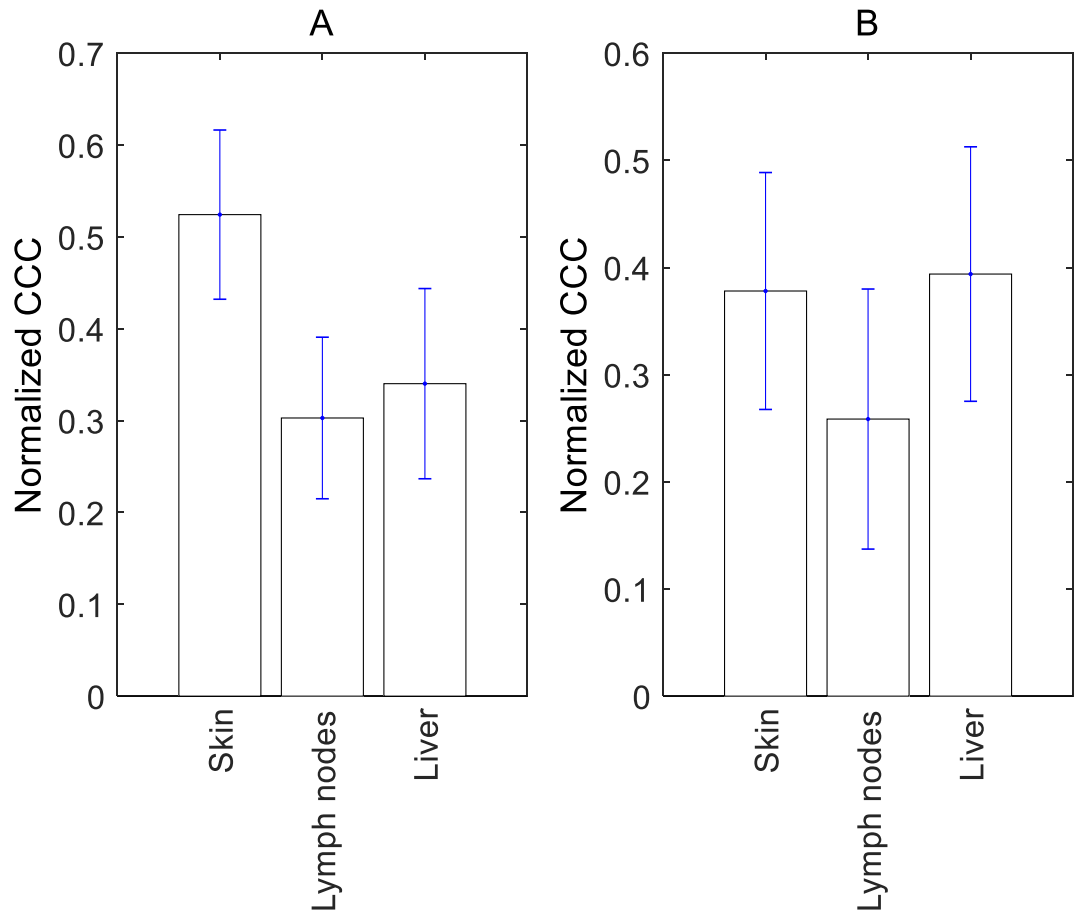

Figure 7 CCC for tumor samples from breast cancer metastases to different organs [50]: (A) Normalized CCC of the collective dissemination associated gene network for breast cancer metastases to different sites: skin (n = 17), lymph nodes (n = 39), and liver (n = 16). (B) Normalized CCC of the IBC associated gene network for breast cancer metastases to different sites as in (A). Error bars indicate the standard error in the estimate of $CCC_{norm}$ calculated using the bootstrap method.

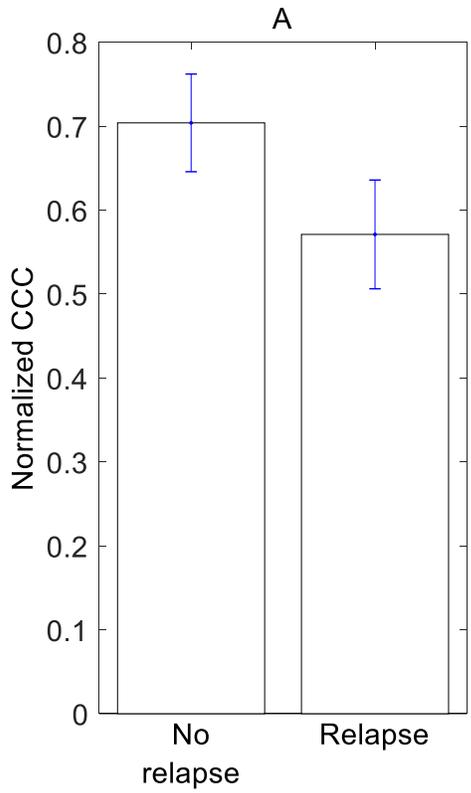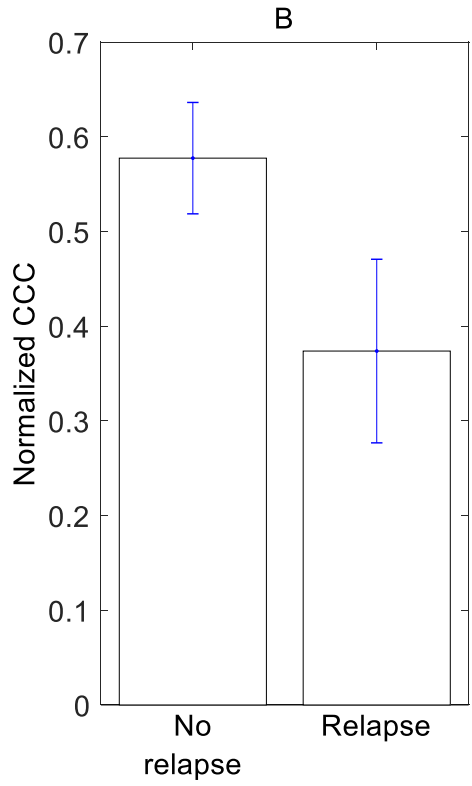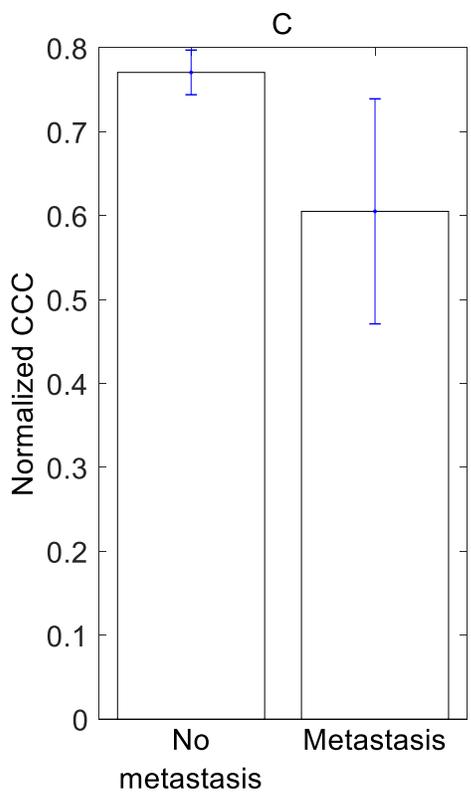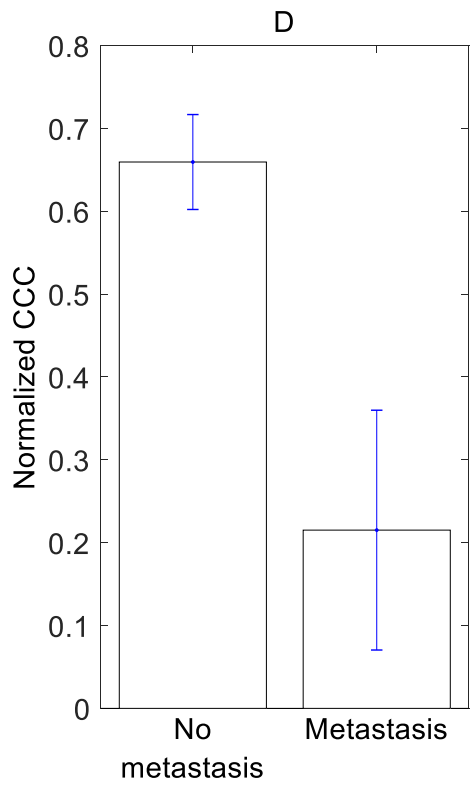

Figure 8 For tumor samples from breast cancer patients who exhibit metastatic relapse during the follow up period (n = 179) and those who do not exhibit breast cancer relapse (n = 107), from the study by Wang *et al.* [47], normalized CCC for (A) collective dissemination associated gene network and (B) IBC associated gene network. Similarly, for breast cancer patient data from The Cancer Genome Atlas (TCGA) [51] (n = 13 for the metastasis group, n = 527 for the non-metastatic group), normalized CCC for (C) collective dissemination associated gene network and (D) IBC associated gene network. Error bars indicate the standard error in the estimate of $CCC_{norm}$ calculated using the bootstrap method.

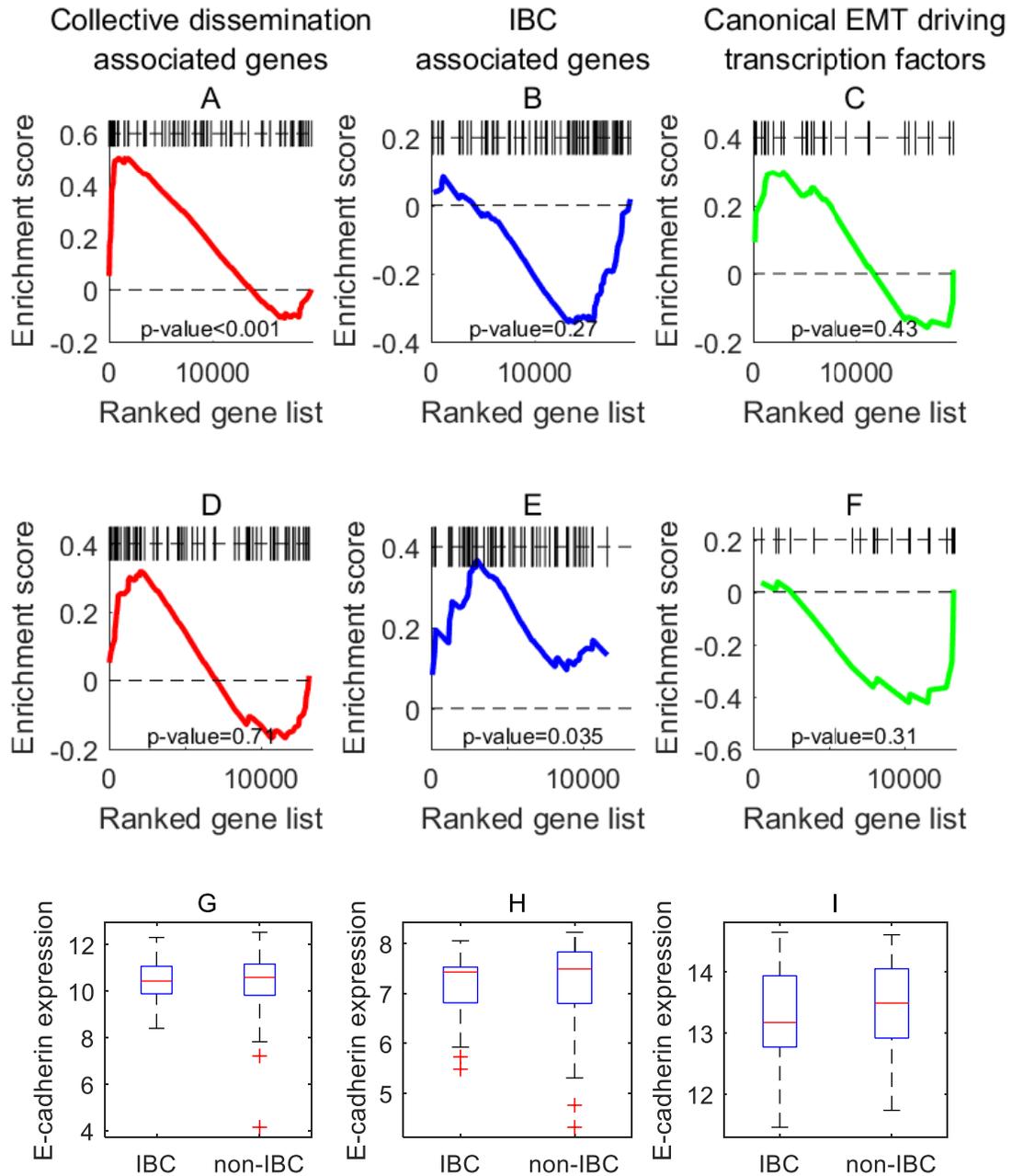

Figure 9 Top two rows (A-F): enrichment profiles for different gene sets in data from Grosse-Wilde *et al.* [36] (top row) and in data from Iwamoto *et al.* [40] (middle row). Black bars along the top of each plot indicate the positions of hits to the gene set along the ordered list of genes ranked by correlation to the phenotype. In the top row of plots, there is high correlation with the epithelial phenotype on the left and it decreases towards right, epithelial versus mesenchymal cell lines GSEA. In the middle row of plots, there is high correlation with the IBC phenotype on the left and correlation decreases towards right, IBC versus non-IBC breast tumor samples GSEA. Nominal p-values of enrichment are indicated at the bottom of each plot. Bottom row (G-I): Mean expression of E-cadherin (CDH1 gene) in tumor samples from IBC and non-IBC patients in studies by (G) Iwamoto *et al.* [40], (H) Boersma *et al.* [41], and (I) Woodward *et al.* [21]. There is no significant difference in the expression of CDH1 gene in any of the three breast cancer datasets. This result indicates that the information provided by the CCC about differences in gene expression between tumor samples from IBC and non-IBC patients cannot be obtained by studying the differential expression of CDH1, a gene which has previously been shown to be associated with IBC [53, 54].

# Supplementary figures

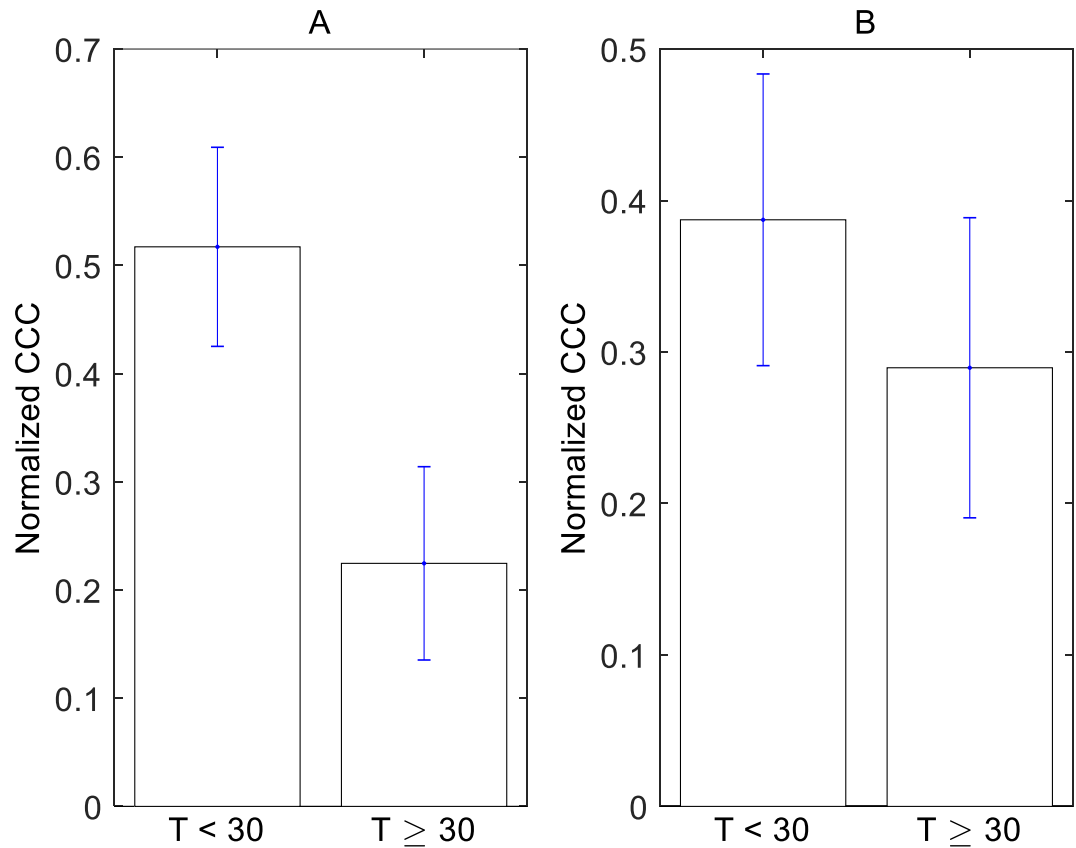

Figure S1 Normalized CCC for estrogen-receptor-positive (ER+) non-IBC breast cancer patients with metastatic relapse within a 30-month period post-treatment (T < 30; n = 39) and with metastatic relapse between 30 and 60 months post-treatment (T ≥ 30; n = 41). Data from the study by Wang *et al.* [47]. (A) Normalized CCC of the collective dissemination associated genes. (B) Normalized CCC of the IBC associated genes. Error bars indicate the standard error in the estimate of $CCC_{norm}$ calculated using the bootstrap method. There were too few estrogen-receptor-negative (ER-) patients in the data set for similar analysis. The trend here is similar to the trend in fig. 7 (A).

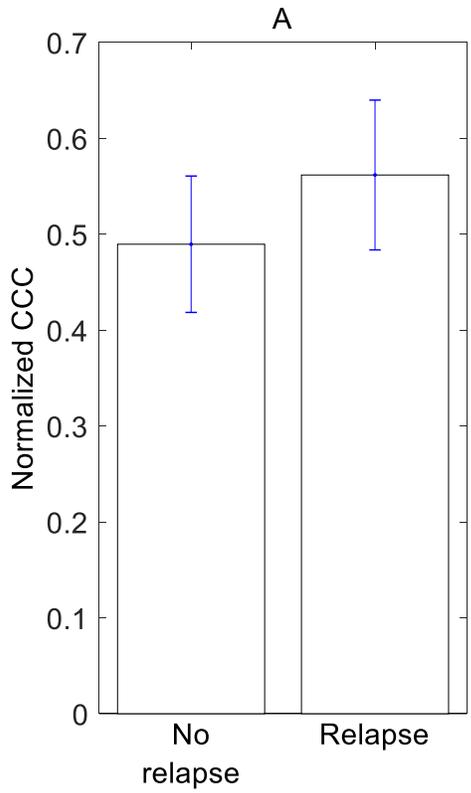
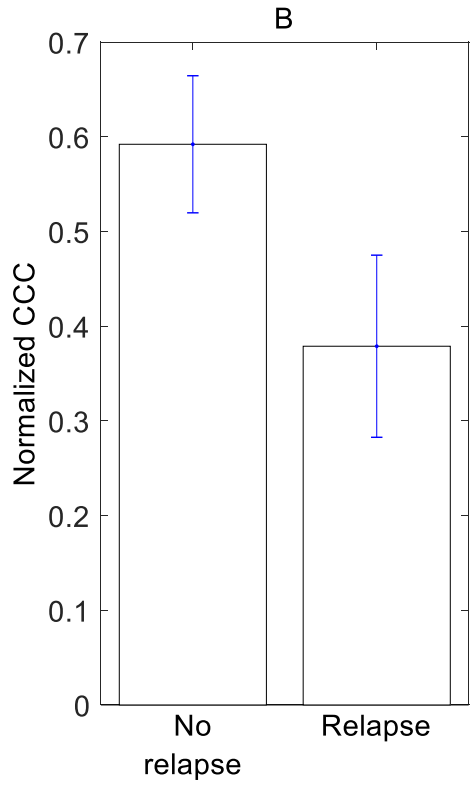
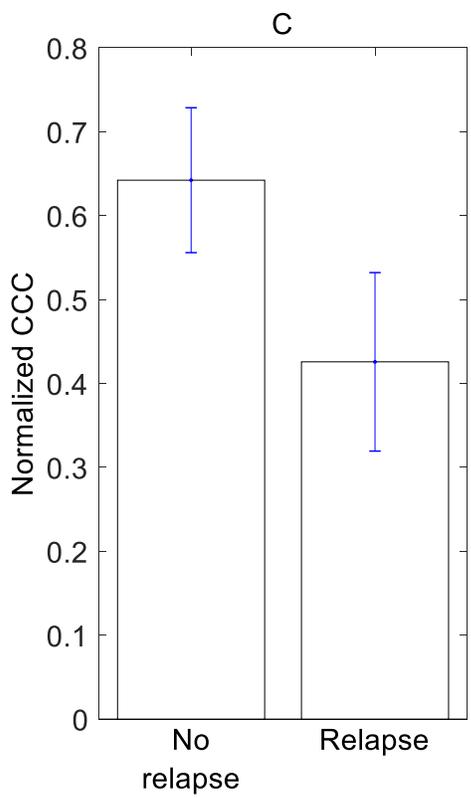
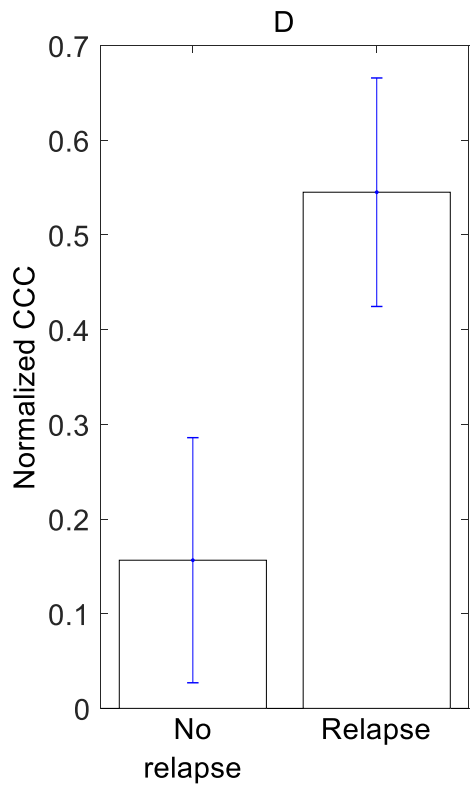

Figure S2 Normalized CCC for breast cancer patients with different estrogen-receptor status, data from Wang *et al.* [47]. Top panel: patients with estrogen-receptor-positive status. This group has 129 patients with no relapse during the 5-year follow-up period and 80 patients with metastatic relapse within 5 years post-treatment. (A) Normalized CCC for the collective dissemination associated genes. (B) Normalized CCC for the IBC associated genes. Bottom panel: patients with estrogen-receptor-negative status. This group has 50 patients with no relapse during the 5-year follow-up period and 27 patients with metastatic relapse within 5 years post-treatment. (A) Normalized CCC for the collective dissemination associated genes. (B) Normalized CCC for the IBC associated genes. Error bars indicate the standard error in the estimate of $CCC_{norm}$ calculated using the bootstrap method.

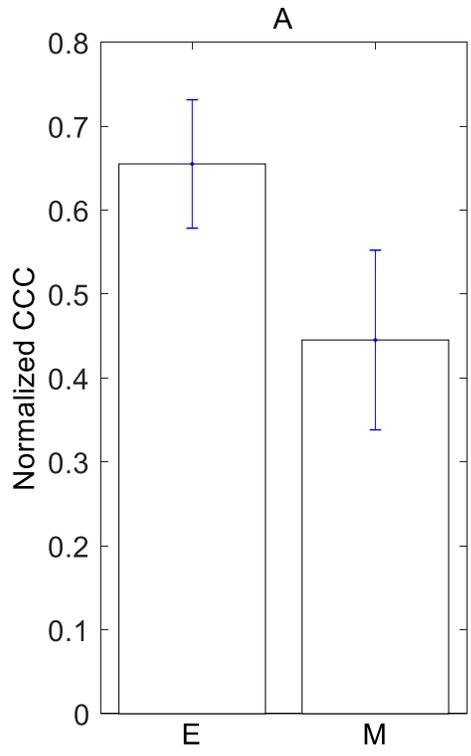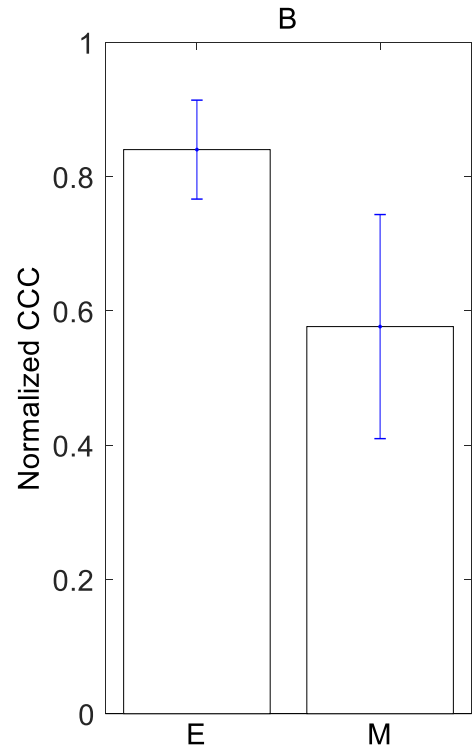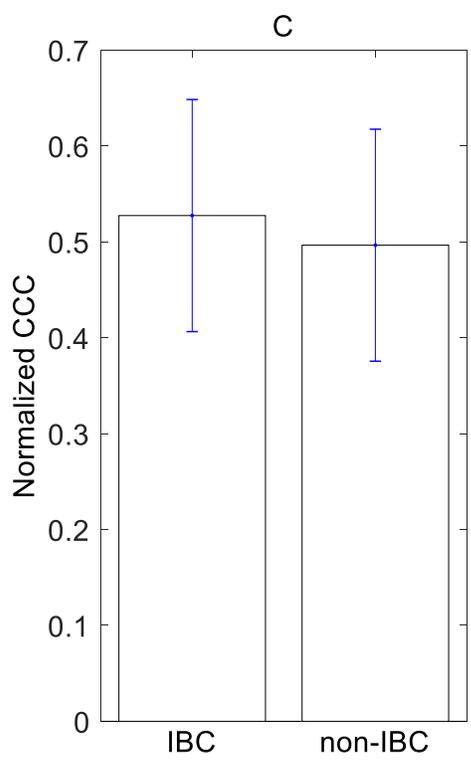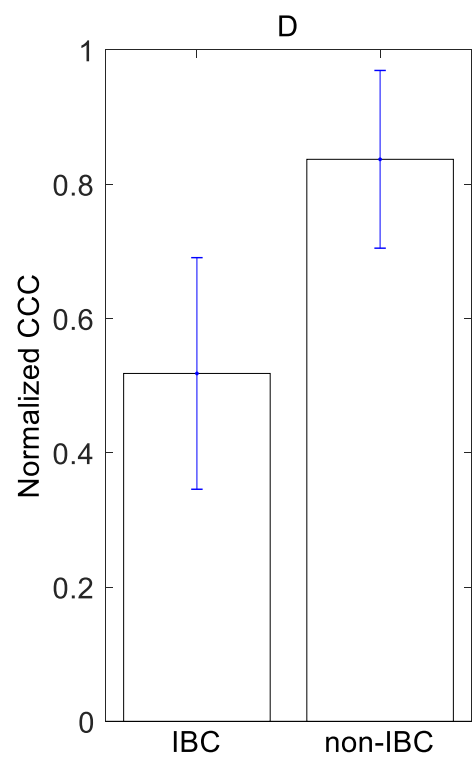

Figure S3 Top panel: Normalized CCC for epithelial (E) (n = 13) and mesenchymal (M) (n = 11) cell lines from the study by Grosse-Wilde *et al.* [36]. (A) Normalized CCC for genes up-regulated in cells in circulating tumor cell clusters [10]. (B) Normalized CCC for genes down-regulated in cells in circulating tumor cell clusters [10]. Bottom panel: Normalized CCC for tumor samples from IBC patients (n = 25) and from non-IBC breast cancer patients (n = 57) from the study by Iwamoto *et al.* [40]. (A) Normalized CCC for genes up-regulated in cells in circulating tumor cell clusters [10]. (B) Normalized CCC for genes down-regulated in cells in circulating tumor cell clusters [10]. Error bars indicate the standard error in the estimate of $CCC_{norm}$ calculated using the bootstrap method. In epithelial cell lines, the CCC is higher for genes that are up-regulated and for genes that are down-regulated in cells in circulating tumor cell clusters, further indicating that the CCC contributes information independent of the levels of gene expression. In tumor samples from breast cancer patients, there is no significant difference in CCC for up-regulated genes between IBC and non-IBC patient groups, while the trend in CCC values for down-regulated genes in opposite to the trend in fig. 2 (A). Taken together, these results indicate that collective consideration of both up-regulated and down-regulated genes is important for understanding the principles underlying phenotypic differences between groups.

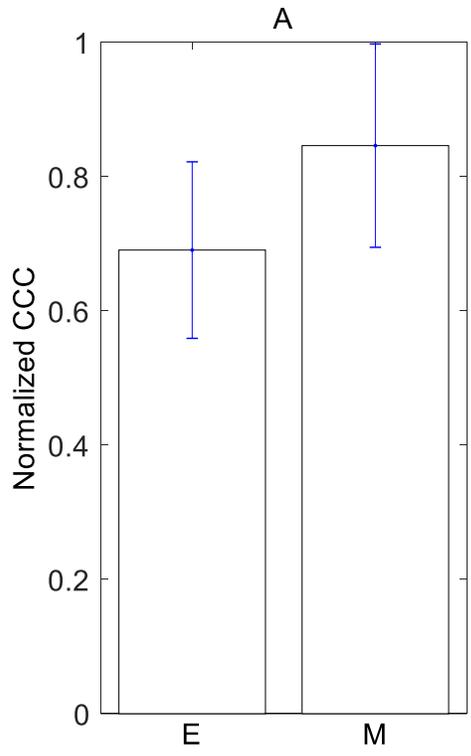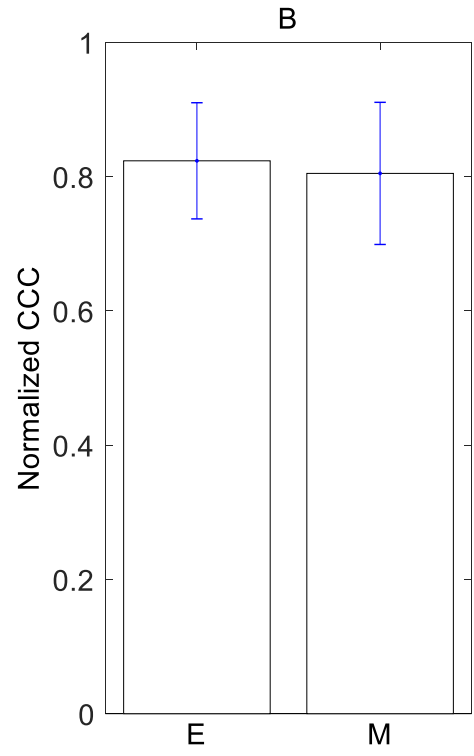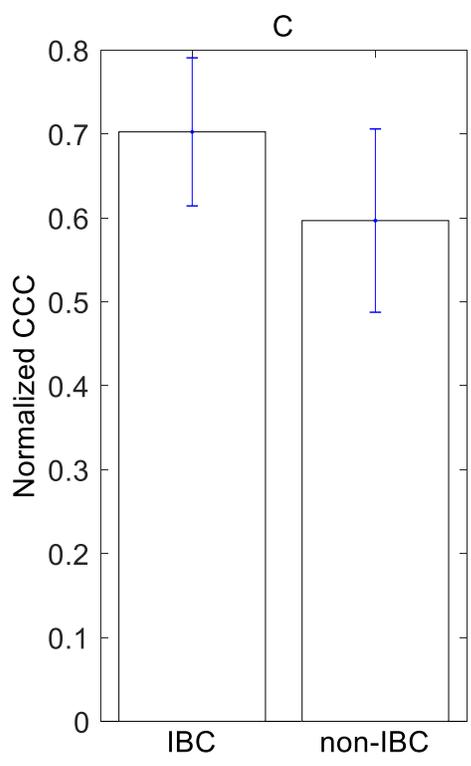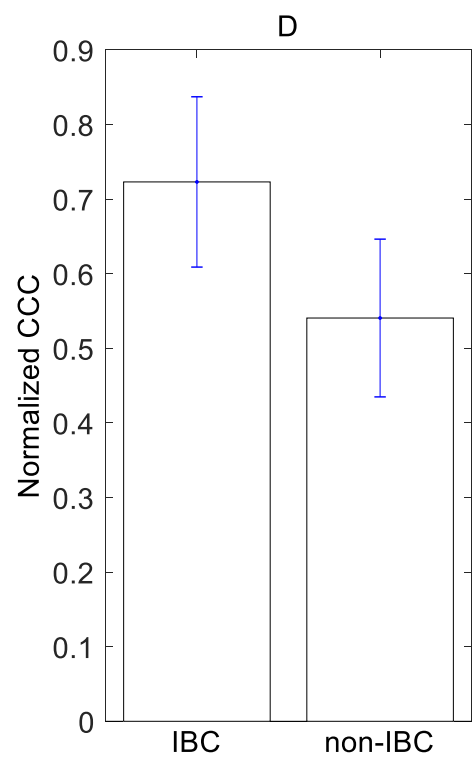

Figure S4 Top panel: Normalized CCC for epithelial (E) (n = 13) and mesenchymal (M) (n = 11) cell lines from the study by Grosse-Wilde *et al.* [36]. (A) Normalized CCC for top 50 genes specifically expressed in epithelial cells [36]. (B) Normalized CCC for top 50 genes specifically expressed in mesenchymal cells [36]. Bottom panel: Normalized CCC for tumor samples from IBC patients (n = 25) and from non-IBC breast cancer patients (n = 57) from the study by Iwamoto *et al.* [40]. (A) Normalized CCC for top 50 genes specifically expressed in epithelial cells [36]. (B) Normalized CCC for top 50 genes specifically expressed in mesenchymal cells [36]. CCCs for both gene sets do not differ significantly between epithelial and mesenchymal cell lines, once again indicating the independence of CCC from gene expression levels and related metrics. Comparing CCCs for both gene sets between IBC and non-IBC patient groups, there is no significant difference, indicating that the IBC phenotype may not directly correlate with epithelial cell associated genes or with mesenchymal cell associated genes.